\documentclass[twocolumn,showpacs,preprintnumbers,amsmath,amssymb]{revtex4-1}

\usepackage{lipsum}
\usepackage{tabularx}
\usepackage{array}
\usepackage{mathrsfs}
\usepackage{amssymb}
\usepackage{amsmath}
\usepackage{cases}  
\usepackage{graphicx}
\usepackage{subfigure} 
\usepackage{dcolumn}
\usepackage{bm}
\usepackage{verbatim} 
\usepackage[dvipdfm,pdfstartview=FitH,colorlinks,linkcolor=blue,anchorcolor=blue,citecolor=blue]{hyperref}
\usepackage{amsthm}  

\begin{document}

\title{Characterizing high-quality high-dimensional quantum key distribution by state mapping between different degree of freedoms}

\author{Fang-Xiang Wang}
\author{Wei Chen}
\email{weich@ustc.edu.cn}
\author{Zhen-Qiang Yin}
\email{yinzq@ustc.edu.cn}
\author{Shuang Wang}
\author{Guang-Can Guo}
\author{Zheng-Fu Han}
\affiliation{${}^1$CAS Key Laboratory of Quantum Information, University of Science and Technology of China, Hefei 230026, China\\
${}^2$Synergetic Innovation Center of Quantum Information $\&$ Quantum Physics, University of Science and Technology of China, Hefei, Anhui 230026, China\\
${}^3$State Key Laboratory of Cryptology, P. O. Box 5159, Beijing 100878, China
}


\begin{abstract}

Quantum key distribution (QKD) guarantees the secure communication between legitimate parties with quantum mechanics. High-dimensional QKD (HDQKD) not only increases the secret key rate but also tolerates higher quantum bit error rate (QBER). Many HDQKD experiments have been realized by utilizing orbital-angular-momentum (OAM) photons as the degree of freedom (DOF) of OAM of the photon is a prospective resource for HD quantum information. In this work we proposed and characterized that a high-quality HDQKD based on polarization-OAM hybrid states can be realized by utilizing state mapping between different DOFs. Both the preparation and measurement procedures of the proof-of-principle verification experiment are simple and stable. Our experiment verified that $(0.60\pm 0.06)\%$ QBER and $1.849\pm 0.008$ bits secret key rate per sifted signal can be achieved for a four-dimensional QKD with the weak coherent light source and decoy state method.

\end{abstract}



\maketitle


Higher-order Poincar\'{e} (HOP) has recently been introduced to describe the total angular momentum of a photon \cite{Milione2011,Milione2012}. Photons with orbital angular momentum (OAM) \cite{Allen1992} are also called twisted photons \cite{Erhard2018}. The poles of the HOP are scalar vortex states and can be expressed as product of polarization and OAM. The polarization of scalar vortex is spatially homogeneous. The states on the HOP except the poles are nonseparable polarization-OAM hybrid states and are named vector vortices \cite{Hall1996,Maurer2007,Zhan2009,Holleczek2011,Naidoo2016}, which are spatially polarized. The Poincar\'{e} sphere, usually being used to describe polarization states, has also been introduced to describe the degree of freedom (DOF) of OAM, where the basis states are two orthogonal OAM states\cite{Padgett1999}. The Hilbert space of OAM is infinite theoretically and photons with more than 10,000 quanta OAM have been realized in experiment \cite{Fickler2016}. Thus, the DOF of OAM of the photon is a prospective high-dimensional (HD) resource for classical optical communication \cite{Wangjian2012,Bozinovic2013,Willner2015,Milione2015,Gregg2015,Zhu2017,Wanga2018} and quantum information \cite{Yao2011,Erhard2018}, such as HD quantum entanglement \cite{Fickler2012}, quantum memory \cite{Ding2016}, quantum teleportation\cite{Wang2015} and quantum key distribution (QKD)\cite{D'Ambrosio2012,Vallone2014}. QKD \cite{Bennett1984} guarantees the secure communication of two parties with quantum physics \cite{Scarani2009}. The traditional QKD protocols, distributing one bit raw key per effective detection issue, develops rapidly in the last decades \cite{Wang2005,Lo2005,Lo2012,Sasaki2014,Lucamarini2018}. HDQKD not only increases the secret key distribution rate but also makes the legitimate parties have more advantages to detect an eavesdropper than the traditional 2-dimensional QKD does\cite{Cerf2002,Sheridan2010,Bradler2016}. Hence, HD quantum key distribution (HDQKD) with OAM photons have been a hot topic in recent years \cite{Nape2018a,Bouchard2018a,Cozzolino2018,Bouchard2018b,Larocque2017,Sit2017,Ndagano2017,Mirhosseini2015,Mafu2013,Groblacher2006}.

On the other hand, polarization is one of the most mature DOF in classical and quantum communication. By combining both Poincar\'{e} spheres of polarization and OAM, the HOP is able to describe the polarization-OAM hybrid vector vortices. Owing to the nonseparable characteristic, the polarization and OAM of the vector vortex state on the HOP can be manipulated by each other. The nonseparable characteristic also reveals the unique feature of the spatially polarized structure of the vector vortex \cite{Hall1996,Maurer2007,Zhan2009,Holleczek2011}.

In this work, by using of the nonseparable feature of the vector vortex, we realized the state mapping between the DOF of OAM and polarization. Then, a proof-of-principle characterization experiment was implemented to verify the feasibility of a four dimensional QKD. By taking advantage of the state mapping method between different DOFs, only polarization manipulation to a vector vortex state on the HOP is necessary to implement the HDQKD. Thus, the preparation and measurement fidelities are determined by the polarization control procedure, which is high-precision. We characterized that a HDQKD system based on our protocol could achieve $1.849\pm 0.008$ bits secret key rate per sifted signal (SKRPSS) with ultra-low quantum-bit-error-rate (QBER) $0.60\%\pm 0.06\%$ and ultra-high stability.


\begin{figure}
	\centering
	\resizebox{8.5cm}{2.83cm}{\includegraphics{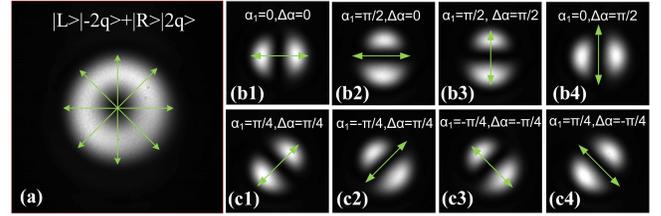}}
	\caption{The experimental realized beam profiles of (a) the radial vector vortex and the corresponding MUBs (b1)-(b4) $\{|\psi\rangle_i\}$ and (c1)-(c4) $\{|\phi\rangle_i\}$ constructed in Equation \ref{eq:unbiasedbasis}. The beam profiles are prepared by a q-plate with $q=1/2$. The green arrows are the orientation of the linear polarization.}
	\label{fig:1}
\end{figure}

\begin{figure}
	\centering
	\resizebox{8cm}{3.45cm}{\includegraphics{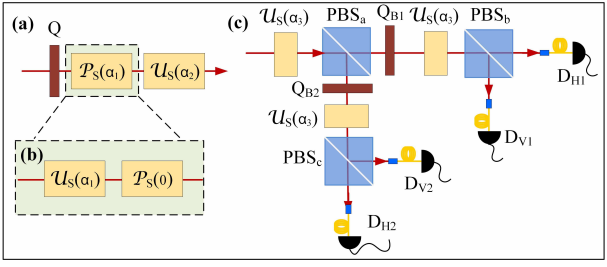}}
	\caption{The schematic setups of preparation and projective measurement of two sets of four dimensional MUBs. (a) The preparation procedure; (b) the factorization of $\hat{\mathcal{P}}_s$; (c) the projective measurement setup. Q, Q${}_{B1(B2)}$: q-plate; D${}_{H(V)}$: single-photon detector.}
	\label{fig:2}
\end{figure}

The vector vortex state on the equator of the HOP is in the form of $|L\rangle|{-l}\rangle+e^{i\theta}|R\rangle|l\rangle$ and is completely spatially polarized, where $|L\rangle$ ($|R\rangle$) is the left (right) circular polarization and $|l\rangle$ is the OAM with $l\hbar$ quantum number. Such a vector vortex can be prepared by the Sagnac loop structure with a spiral phase plate\cite{Milione2012,Rafsanjani2015} or a q-plate \cite{Marrucci2006,Nagali2009,Marrucci2011}, among which the q-plate method is easier to realize \cite{Marrucci2011,D'Ambrosio2012,D'Ambrosio2013}. A q-plate can be expressed as $\hat{\mathcal{Q}}(q)=(|L\rangle\langle R|)\otimes(\sum_{l}{|{l-2q}\rangle}\langle l|)+(|R\rangle\langle L|)\otimes(\sum_{l}{|{l+2q}\rangle}\langle l|)$, where $2q$ is the quantum number of OAM and is determined by the q-plate. As the Hilbert space of OAM is not closed, the summation of the q-plate is also not closed, in principle. For a linear polarized input Gaussian beam, the output state from the q-plate is $|\mathcal{V}\rangle=\hat{\mathcal{Q}}(q)|H\rangle|0\rangle=(1/\sqrt{2})(|R\rangle|2q\rangle+|L\rangle|-2q\rangle)$. It is a radial polarized vector vortex (see Figure \ref{fig:1}(a)) on the equator of the HOP. If we project the state into homogeneous linear polarization, then $\hat{\mathcal{P}}_s(\alpha)|\mathcal{V}\rangle=(e^{-i\alpha}/2)(cos\alpha|H\rangle+sin\alpha|V\rangle)(|2q\rangle+e^{2i\alpha}|-2q\rangle)$, where $\hat{\mathcal{P}}_s(\alpha)=(cos\alpha|H\rangle+sin\alpha|V\rangle)(cos\alpha\langle H|+sin\alpha\langle V|)$ is the projector, $\alpha$ is the projection angle and the subscript $s$ means that the manipulation is operated on the DOF of polarization. The projective state becomes a product state of polarization and OAM. Figure \ref{fig:1}(a) gives the beam profiles of the radial polarized vector vortex prepared by a q-plate with ${q=1}$. By rotating the projection angle $\alpha$, the projective polarization state rotates on the equator of the polarization Poincar\'{e} sphere and the projective OAM state rotates on the equator of the OAM Poincar\'{e} sphere \cite{Padgett1999} simultaneously, though we have not manipulated the state in the DOF of OAM directly. Thus, we only need operating on a single DOF to fulfill the manipulation of multiple DOF. For example, by applying another unitary transform $\hat{\mathcal{U}}_s(\alpha)$ in the DOF of polarization after the projector, the state becomes $\hat{\mathcal{U}}_s(\alpha_2)\hat{\mathcal{P}}_s(\alpha_1)|\mathcal{V}\rangle=(e^{-i\alpha_1}/2)\big(cos(\Delta\alpha)|H\rangle+sin(\Delta\alpha)|V\rangle\big)\big(|2q\rangle+e^{2i\alpha_1}|{-2q}\rangle\big)$, where $\Delta\alpha=\alpha_2-\alpha_1$ and the unitary transform $\hat{\mathcal{U}}_s(\alpha)$ is given by 
\begin{displaymath}
\hat{\mathcal{U}}_s(\alpha)=
\begin{pmatrix}
cos\alpha & sin\alpha\\
sin\alpha & -cos\alpha
\end{pmatrix}.
\end{displaymath}
Then, by choosing $\alpha_1,\Delta\alpha\in\{0,{\pi/2}\}$ and $\alpha_1,\Delta\alpha\in\{\pi/4,{-\pi/4}\}$, two sets of four dimensional MUBs $\{|\psi\rangle_i\}$ and $\{|\phi\rangle_i\}$ are constructed, respectively. $\{|\psi\rangle_i\}$ and $\{|\phi\rangle_i\}$ are defined as
\begin{equation}
\begin{aligned}
\big\{&|\psi\rangle_1=|H\rangle(|2q\rangle+{|{-2q}\rangle}),|\psi\rangle_2=|H\rangle(|2q\rangle-|{-2q}\rangle),\\
&|\psi\rangle_3=|V\rangle(|2q\rangle-|{-2q}\rangle),|\psi\rangle_4=|V\rangle(|2q\rangle+|{-2q}\rangle)\big\},\\
\big\{&|\phi\rangle_1=|D\rangle(|2q\rangle+i|{-2q}\rangle),|\phi\rangle_2=|D\rangle(|2q\rangle-i|{-2q}\rangle),\\
&|\phi\rangle_3=|A\rangle(|2q\rangle-i|{-2q}\rangle),|\phi\rangle_4=|A\rangle(|2q\rangle+i|{-2q}\rangle)\big\}
\end{aligned}
\label{eq:unbiasedbasis}
\end{equation}
where $|D(A)\rangle=\frac{1}{\sqrt{2}}(|H\rangle\pm|V\rangle)$. Figures \ref{fig:1}(b1)-(b4) and Figures \ref{fig:1}(c1)-(c4) give the beam profiles of $\{|\psi\rangle_i\}$ and $\{|\phi\rangle_i\}$, respectively. Only polarization control is necessary for the MUBs preparation procedure. The MUBs preparation procedure (Figure \ref{fig:2}(a)) can be realized with high-speed devices because $\hat{\mathcal{P}_s}(\alpha_1)$ can be factorized by $\hat{\mathcal{P}_s}(\alpha_1)=\hat{\mathcal{U}}_s(\alpha_1)\hat{\mathcal{P}_s}(0)$ (see Figure \ref{fig:2}(b)). By applying this factorization, only two active unitary transforms $\hat{\mathcal{U}}_s(\alpha_1)$ and $\hat{\mathcal{U}}_s(\alpha_2)$ are necessary for the MUBs preparation procedure. The MUBs can then be generated with high speed by realizing $\hat{\mathcal{U}}_s(\alpha_1)$ and $\hat{\mathcal{U}}_s(\alpha_2)$ with electro-optical modulators (EOMs). Then, the MUBs in Equation \ref{eq:unbiasedbasis} can be constructed by choosing $\alpha_1,\alpha_2\in\{0,{\pi/2}\}$ and $\alpha_1,\alpha_2\in\{\pi/4,{-\pi/4}\}$, respectively.

Above, we have proposed the preparation of two sets of four dimensional MUBs by utilizing the nonseparable feature of vector vortices. Another important part of a QKD system is the complete projective measurement of these MUBs. Now we construct the complete projective measurement setup as below. 

A q-plate can manipulate both DOFs of polarization and OAM (as shown by $\hat{\mathcal{Q}}(q)$). According to this property of the q-plate, we propose a simple setup to convert the OAM projection to polarization projection (see Figure \ref{fig:2}(c)) and realize the high-precision complete projective measurement for the prepared MUBs. In Figure \ref{fig:2}(c), the unitary transform $\hat{\mathcal{U}}_s(\alpha_3)$ is used to choose projective basis from $\mathcal{M}_1=\sum_{i}|\psi\rangle_i\langle\psi|$ ($\alpha_3=0$) and $\mathcal{M}_2=\sum_{i}|\phi\rangle_i\langle\phi|$ ($\alpha_3=\pi/4$). Without the loss of generality, we take $\mathcal{M}_1$ as an example. $\hat{\mathcal{U}}_s(\alpha_3)$ becomes a unit matrix. The first polarizing beam splitter (PBSa) realizes the projective measurement of polarization. Then two q-plates (Q${}_{B1}$ and Q${}_{B2}$) are used to twist the polarization and OAM:
\begin{equation}
\begin{aligned}
&\hat{\mathcal{Q}}_{B1}|H\rangle(|{2q}\rangle+e^{2i\alpha_1}|{-2q}\rangle)\\
=&{(1/\sqrt{2})}[(|L\rangle+e^{2i\alpha_1}|R\rangle)|0\rangle+|R\rangle|{4q}\rangle+e^{2i\alpha_1}|L\rangle|{-4q}\rangle],\\
&\hat{\mathcal{Q}}_{B2}|V\rangle(|{2q}\rangle+e^{2i\alpha_1}|{-2q}\rangle)\\
=&{(1/\sqrt{2})}[(-|L\rangle+e^{2i\alpha_1}|R\rangle)|0\rangle+|R\rangle|{4q}\rangle-e^{2i\alpha_1}|L\rangle|{-4q}\rangle]
\end{aligned}
\label{eq:QB1QB2}
\end{equation} 
According to Equation \ref{eq:QB1QB2}, the input OAM state $|{2q}\rangle+e^{2i\alpha_1}|{-2q}\rangle$ has been mapped to the polarization state $|L\rangle+e^{2i\alpha_1}|R\rangle)$ with zero OAM. By ignoring the non-zero OAM terms, the input state can be discriminated by polarization projection now. Thus, by cascading a PBS (PBSb and PBSc) and a single-mode fiber (SMF) in each path, we project $|\psi\rangle_1$, $|\psi\rangle_2$, $|\psi\rangle_3$ and $|\psi\rangle_4$ to the single-photon detectors (SPDs) D${}_{H1}$, D${}_{V1}$, D${}_{H2}$ and D${}_{V2}$, respectively. While ${|\phi\rangle_i}$ projects to all SPDs with equal probabilities. The SMFs are used to filter out all non-zero OAM terms and thus leads to a 50\% efficiency loss. The similar conclusion can be obtained for the projective basis $\mathcal{M}_2$. Hence, we have realized the complete projective measurement of the two sets of four dimensional MUBs by utilizing the "twist" ability of q-plates. Our projective measurement method possesses high resolution as the extinction ratio of polarization projective measurement is much higher than the directly OAM projective measurement.


\begin{figure}
	\centering
	\resizebox{8cm}{5.0cm}{\includegraphics{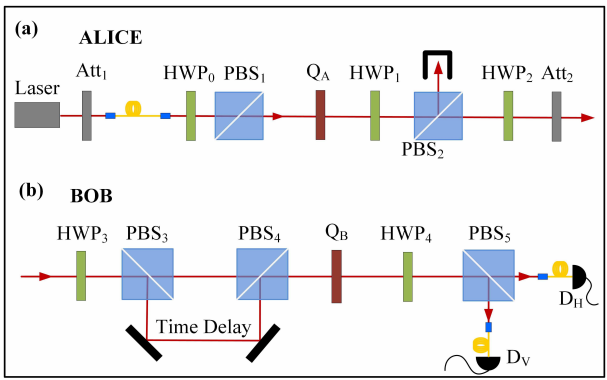}}
	\caption{Experimental setup of the HD-QKD. (a) The state preparation setup of Alice; (b) The projective measurement setup of Bob. Q${}_A$, Q${}_B$: q-plate; Att: optical attenuator.}
	\label{fig:3}
\end{figure}

Without the loss of generality, a proof-of-principle experiment has been implemented to verify the feasibility of a four dimensional BB84-protocol QKD \cite{Bennett1984} (as shown in Figure \ref{fig:3}), where Alice prepares and transmits the states to Bob through the quantum channel and Bob makes projective measurements to the received states. In the state preparation procedure of Alice, the light source is a Ti:sapphire locked pulsed laser. The wavelength of the laser is $(780\pm 5)$nm, the pulse width is within 200fs and the repeating rate is 77MHz. $\hat{\mathcal{U}}_s(\alpha_1)$ and $\hat{\mathcal{U}}_s(\alpha_2)$ are realized by HWPs (HWP${}_1$ and HWP${}_2$ in Figure \ref{fig:3}(a)). The light intensity is attenuated to single photon level by attenuators (Att1 and Att2) before transmitted to Bob. The attenuators are also used to realize the "vacuum+weak decoy state" method \cite{Wang2005,Lo2005,Ma2005}. The average photon numbers of the transmitted signal and decoy states are $\mu\simeq 0.053$ and $\nu\simeq 0.017$, respectively. In the projective measurement procedure of Bob, basis chosen $\hat{\mathcal{U}}_s(\alpha_3)$ is realized by HWPs (HWP${}_3$ and HWP${}_4$ in Figure \ref{fig:3}(b)): the rotation angles of HWP${}_3$ and HWP${}_4$ are zero for $\mathcal{M}_1$ and are $\pi/8$ for $\mathcal{M}_2$, respectively. A 3.05ns time delay is used to multiplexing the transmissive and reflective signals of PBS3 to the following identical path after PBS4. SMFs are used to project the zero OAM terms into single-photon detectors (SPDs, SPCM-AQRH-14 of Excelitas Technologies Corp.). The detection efficiency and dark count rate of the SPD are 60\% and 67Hz, respectively. The time resolution of the SPD is 350ps with full width at half maximum (FWHM). 

\begin{figure}
	\centering
	\resizebox{8cm}{4.17cm}{\includegraphics{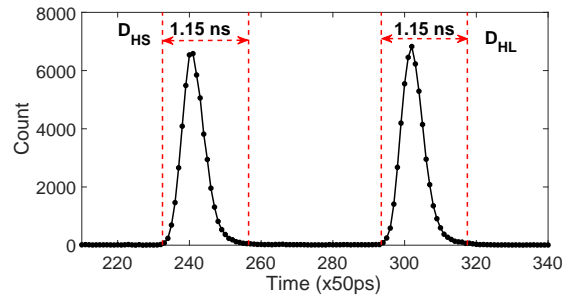}}
	\caption{The arrival time distribution of the SPD D${}_{H}$ in Figure \ref{fig:3}(b).}
	\label{fig:4}
\end{figure}

We use a time-to-digital converter (TDC) with 50ps time resolution to detect the arrival time of photons through the short and long paths of Bob. Figure \ref{fig:4} gives the detection results of D${}_H$ in Figure \ref{fig:3}(b). According to the time distribution of count events, the effective detection window of the SPDs are set to 1.15ns with post-selection by the TDC. Figure \ref{fig:4} shows that 3.05ns time delay is sufficient for time resolution between the short and long paths. The effective dark count rate for the post-selection detection window is then $7.7\times 10^{-8}$ per pulse.

\begin{figure}
	\centering
	\resizebox{8cm}{5.63cm}{\includegraphics{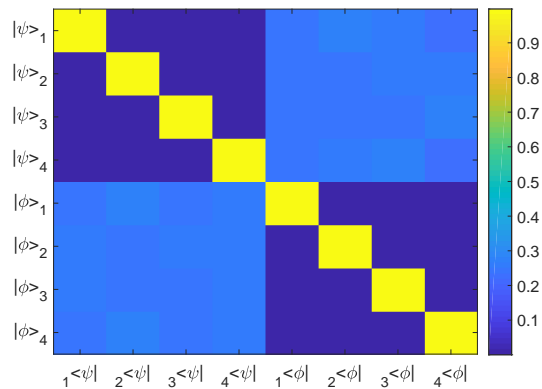}}
	\caption{The probability distribution of the projective measurement setup for transmitted signal states with $\mu\simeq 0.053$.}
	\label{fig:5}
\end{figure}

\begin{table}
	\centering
	\scriptsize
	\caption{The projective efficiency $\mathcal{E}$ and measurement error $3\sigma$ (3 times of standard deviation) of the projective measurement setup.}
	\begin{tabular}{ccccccccc}
		\hline
		& $|\psi\rangle_1$ & $|\psi\rangle_2$ & $|\psi\rangle_3$ & $|\psi\rangle_4$ & $|\phi\rangle_1$ & $|\phi\rangle_2$ & $|\phi\rangle_3$ & $|\phi\rangle_4$\\
		\hline
		$\mathcal{E}_\mu$ & 99.35\% & 99.32\% & 99.42\% & 99.68\% & 99.19\% & 99.73\% & 99.16\% & 99.33\%\\
		$3\sigma_\mu$ & 0.05\% & 0.06\% & 0.05\% &0.04\% & 0.06\% & 0.04\% & 0.07\% & 0.06\%\\
		$\mathcal{E}_\nu$ & 99.27\% & 99.24\% & 99.41\% & 99.67\% & 99.13\% & 99.64\% & 99.12\% & 99.31\%\\
		$3\sigma_\nu$ & 0.10\% & 0.10\% & 0.09\% &0.07\% & 0.11\% & 0.08\% & 0.13\% & 0.10\%\\
		\hline 
	\end{tabular}
	\label{tab:projectiveefficiency}
\end{table}

The projective measurement results of Bob's setup for transmitted signal states are shown in Figure \ref{fig:5}. The results show very low crosstalk as all diagonal elements are larger than 0.99. We define projective efficiency $\mathcal{E}=1-e_b$ to evaluate the projective quality, where $e_b$ is the QBER of the projective measurement of a transmitted state. For example, $e_b=\sum_{i\neq 1}{}_i\langle\psi|\psi\rangle_1$ for $|\psi\rangle_1$. The projective efficiencies of signal and decoy states are shown in Table \ref{tab:projectiveefficiency}. The projective efficiencies are all larger than $99\%$. The maximum measurement errors are as less as ${3\sigma_\mu}_{max}\simeq 0.07\%$ and ${3\sigma_\nu}_{max}\simeq 0.10\%$ for signal and decoy states, respectively. According to Table \ref{tab:projectiveefficiency}, the average QBERs of the signal and decoy states are $e_{\mu}\simeq (0.60\pm 0.06)\%$ and $e_{\nu}\simeq (0.65\pm 0.10)\%$, respectively.

According to the GLLP formula \cite{Gottesman2004,Lo2005} for HDQKD \cite{Cerf2002}, the secret key rate of the d-dimensional BB84-QKD can be estimated by
\begin{equation}
\mathcal{R}\geq {q_m}Q_\mu\{-f_{EC}(e_\mu)H_d(e_\mu)+\Delta_1[log_2d-H_d(e_1)]\}
\label{siftkeyrate}
\end{equation}
where ${q_m}$ depends on the QKD protocol and is $1/2$ in our experiment, $Q_\mu$ and $e_\mu$ are the gain and QBER of the signal state, respectively, $\Delta_1$ is the fraction of single-photon signals transmitted by Alice, $f_{EC}(e_\mu)$ is the error correction efficiency of the signal state and $H_d(e_\mu)=-(1-e_\mu)log_2(1-e_\mu)-e_\mu log_2(\frac{e_\mu}{d-1})$ is the d-dimensional Shannon entropy. By applying the "vacuum+weak decoy state" method \cite{Ma2005}, $\delta_1$ and $e_1$ can be estimated by the following constraint inequalities: $\Delta_1\geq\frac{\mu^2e^{-\mu}}{\mu\nu-{\nu}^2}\big(\frac{Q_\nu}{Q_\mu} e^\nu-\frac{{\nu}^2}{\mu^2}e^\mu-\frac{\mu^2-\nu^2}{\mu^2}\frac{Y_0}{Q_\mu}\big)$, $Y_1^{L,\nu,0}=\frac{\mu}{\mu\nu-{\nu}^2}\big(Q_\nu e^\nu-\frac{{\nu}^2}{{\mu}^2}Q_\mu e^\mu-\frac{{\mu}^2-{\nu}^2}{{\mu}^2}Y_0\big)$ and $e_1\leq \frac{e_\nu Q_\nu e^{\nu}-e_0Y_0}{Y_1^{L,\nu,0}\nu}$, where $Y_0$ is the dark count rate of the SPD, $e_0$ is the QBER of the vacuum state, $Q_\nu$ and $e_\nu$ are the gain and QBER of the weak decoy state, respectively.
The corresponding parameters of our experiment are $\mu\simeq 0.053, \nu\simeq 0.017, Q_\mu\simeq 4.03\times 10^{-3}, Q_\nu\simeq 1.33\times 10^{-3},e_{\mu}\simeq (0.60\pm 0.06)\%, e_{\nu}\simeq (0.65\pm 0.10)\%, Y_0\simeq 8\times 10^{-8}$ and $e_0=0.5$. Then, we obtain the calculated SKRPSS $\mathcal{R}/({q_m}Q_{\mu})\simeq 1.849\pm 0.008$, where we have set $f_{EC}(e_\mu)=1$ (the SKRPSS is about 1.840 bits by considering a practical error correction efficiency $f_{EC}(e_\mu)=1.15$).


\begin{table*}
	\centering
	\caption{Secret key rate comparison of some typical OAM-based HDQKD works with our works. $d$ is the dimensionality of the QKD system. WCS: weak coherent laser source;"---":not shown in the reference; PPKTP and BBO represent the single-photon sources generated by spontaneous parametric-down-conversion process with periodically poled potassium titanyl phosphate and barium borate nonlinear crystals, respectively.}
	\begin{tabular}{cccccccccccc}
		\hline
		& Our work & Ref. \cite{Nape2018a} & Ref. \cite{Bouchard2018a} & Ref. \cite{Cozzolino2018} & Ref. \cite{Bouchard2018b}  & Ref. \cite{Larocque2017} & Ref. \cite{Sit2017} & Ref. \cite{Ndagano2017} & Ref. \cite{Mirhosseini2015}& Ref. \cite{Mafu2013} & Ref. \cite{Groblacher2006}\\
		\hline
		 $\mathcal{R}/({q_m}Q_\mu)$ & 1.849 & 1.32 & $<1.3$ &---& 1.5316 & 0.49 & 0.39 & 1.63 & 2.05 & 1.139 & 0.7994 \\
		 $e_\mu$ & 0.60\% & 4\% &---& 14.1\% & 3.79\% & 21.2\% & 14\% & 3\% & 10.5\% & 8.8\% & 9.3\%\\
		 $d$ & 4 & 4 & 4 & 4 & 4 & 10 & 4 & 4 & 7 & 4 & 3\\
		 Basis & spin-OAM & spin-OAM & OAM & spin-OAM & OAM & OAM & spin-OAM & spin-OAM & OAM & OAM & OAM \\
		 Source & WCS & PPKTP & BBO & WCS & BBO & PPKTP & PPKTP & WCS & WCS & BBO & BBO\\
		 Channel & Table & Table & Table & 1.2km & Table & Table & 300m & Table & Table &Table & Table\\
		\hline 
	\end{tabular}
	\label{tab:sifted key rate}
\end{table*}

According to the experimental results, the QBER and the corresponding error of our experiment are very low $e_\mu\simeq(0.6\pm 0.06)\%$. This owes to the state preparation and measurement methods proposed here. For both state preparation and measurement procedures, we successfully map OAM superposition state manipulation into polarization manipulation. As polarization manipulation technologies are mature and with high-precision, our protocol achieves very low QBER. Furthermore, the mapping procedures apply no interferometric structure (e.g., the Mach-Zehnder interferometer (MZI)) and makes the system stable (the experimental error is less than 0.1\%). It should be addressed that though the laser repetition rate is 77 MHz, the QBERs of different transmitted states are measured one by one by rotating the HWPs manually in our proof-of-principle experiment. 

A performance comparison of HDQKD systems based on OAM photons is shown in Table \ref{tab:sifted key rate}. where we list some key indexes of HDQKD systems. The QBER of ours is the lowest and thus the corresponding SKRPSS is much higher than that of other four dimensional systems.Though Ref. \cite{Mirhosseini2015} achieved a higher SKRPSS 2.05 bits, it is a seven dimensional QKD system with 10.5\% QBER.

The commonly used tools for OAM state preparation are the spatial light modulator (SLM), the q-plate and the digital micro-mirror device (DMD) \cite{Yao2011,Erhard2018,Marrucci2011}. The main disadvantage of these tools is the relatively low generation speed (from 10 Hz to several kHz \cite{Mirhosseini2015}). By combining with other means such as multi-path encoding method, the generation speed difficulty can be overcome. However, these means are still complex. For example, the high-generation-rate method of Ref. \cite{Cozzolino2018} requires two cascading MZIs. The state generation rate of ours can achieve 100MHz by replacing HWP1 and HWP2 in Figure \ref{fig:3}(a) with free-space EOMs. The free-space EOM could hardly achieve higher modulation rate as the high half-wave voltage required. Though there is still a long way to make them practical, the integrated optical waveguides \cite{Cheny2018} and modulators \cite{Mousavi2017,Mousavi2018} may be a potential solution to high-speed modulation of OAM photons.

The OAM and polarization-OAM state measurement are usually carried out by the hologram-based OAM filter \cite{Yao2011,Gibson2004}, the interferometric method \cite{Leach2002,Leach2004}, the q-plate method with wave plates \cite{Marrucci2011} and the geometric transformation method \cite{Berkhout2010,O'Sullivan2012,Mirhosseini2013,Wen2018}. The hologram-based OAM filter is lack of efficiency. The geometric transformation method can fulfill the projective measurement of tens of OAM modes simultaneously. However, the crosstalk between adjacent 
modes is still a little higher \cite{Mirhosseini2013} for high-quality QKD systems (see Refs. \cite{Larocque2017,Mirhosseini2015,Ndagano2017,Nape2018b,Bouchard2018b} of Table \ref{tab:sifted key rate}). The q-plate method used in previous works is actually the mode filter \cite{Nape2018a,Sit2017,Marrucci2011}. By mapping the OAM projection into polarization projection, we achieve high-precision complete projective measurement of polarization-OAM hybrid states with a much simpler setup. It should be emphasized that only comparing with QBER and SKRPSS of HDQKD systems in Table \ref{tab:sifted key rate} are not rigorous in a sense. Because the photon sources, encoding states and communication channels vary from one to one. However, Table \ref{tab:sifted key rate} gives a clear picture of some key indexes (such as QBER and secret key rate) and shows relative merits of different systems. 


In conclusion, we have proposed and verified the feasibility of a four dimensional QKD protocol by utilizing the nonseparable feature of the vector vortex and the polarization-OAM interconversion ability of the q-plate. The characterization experiment achieves a very low QBER and experimental error. The SKRPSS is as high as $1.849\pm 0.008$ bits with weak coherent light source. Our work enhances the performance quality of HDQKD with polarization-OAM hybrid states. The no-interferometric and precise preparation and measurement methods are ready for practical HDQKD systems. Furthermore, our work also reveals that HD quantum state manipulation can be simplified in some important scenarios by utilizing the nonseparable feature of vector vortex photons.

\section*{Acknowledgments}
\label{acknowledgments}

The authors acknowledge Martin P. J. Lavery for fruitful discussion. This work has been supported by the National Key Research And Development Program of China (Grant Nos. 2016YFA0302600, 2018YFA0306400), the National Natural Science Foundation of China (Grant Nos. 61627820, 61775207, 61675189, 61622506, 61475148, 61575183) and Anhui Initiative in Quantum Information Technologies.


\begin{thebibliography}{99}
\label{thebibliography}

\bibitem{Milione2011} G. Milione, H. I. Sztul, D. A. Nolan, and R. R. Alfano, Higher-Order Poincar\'{e} Sphere, Stokes Parameters, and the Angular Momentum of Light, Phys. Rev. Lett. 107, 053601 (2011).

\bibitem{Milione2012} G. Milione, S. Evans, D. A. Nolan, and R. R. Alfano, Higher Order Pancharatnam-Berry Phase and the Angular Momentum of Light Giovanni, Phys. Rev. Lett. 108, 190401 (2012).

\bibitem{Allen1992} L. Allen, M. W. Beijersbergen, R. J. C. Spreeuw, and J. P. Woerdman, Orbital angular momentum of light and the transformation of Laguerre-Gaussian laser modes, Phys. Rev. A 45, 8185 (1992).

\bibitem{Erhard2018} M. Erhard, R. Fickler, M. Krenn, and A. Zeilinger, Twisted photons: new quantum perspectives in high dimensions, Light Sci. Appl. 7, 17146 (2018).

\bibitem{Hall1996} D. G. Hall, Vector-beam solutions of Maxwell's wave equation Dennis, Opt. Lett. 21, 9 (1996).

\bibitem{Maurer2007} C. Maurer, A. Jesacher, S. F$\ddot{u}$rhapter, S. Bernet, and M. Ritsch-Marte, Tailoring of arbitrary optical vector beams, New J. Phys. 9, 78 (2007).

\bibitem{Zhan2009} Q. Zhan, Cylindrical vector beams: from mathematical concepts to applications, Adv. Opt. Photonics 1, 1 (2009).

\bibitem{Holleczek2011} A. Holleczek, A. Aiello, C. Gabriel, C. Marquardt, and G. Leuchs, Classical and quantum properties of cylindrically polarized states of light, Opt. Express 19, 9714 (2011).

\bibitem{Naidoo2016} D. Naidoo, F. S. Roux, A. Dudley, I. Litvin, B. Piccirillo, L. Marrucci, and A. Forbes, Controlled generation of higher-order Poincar\'{e} sphere beams from a laser, Nat. Photonics 10, 327 (2016).

\bibitem{Padgett1999} M. J. Padgett and J. Courtial, Poincar\'{e}-sphere equivalent for light beams containing orbital angular momentum, Opt. Lett. 24, 430 (1999).

\bibitem{Fickler2016} R. Fickler, G. Campbell, B. Buchler, P. K. Lam, and A. Zeilinger, Twisted light transmission over 143 km, Proc. Natl. Acad. Sci. 113, 13648 (2016).

\bibitem{Wangjian2012} J. Wang, J.-Y. Yang, I. M. Fazal, N. Ahmed, Y. Yan, H. Huang, Y. Ren, Y. Yue, S. Dolinar, M. Tur, and A. E. Willner, Terabit free-space data transmission employing orbital angular momentum multiplexing, Nat. Photonics 6, 488 (2012).

\bibitem{Bozinovic2013} N. Bozinovic, Y. Yue, Y. Ren, M. Tur, P. Kristensen, H. Huang, A. E. Willner, and S. Ramachandran, Terabit-Scale Orbital Angular Momentum Mode Division Multiplexing
in Fibers, Science 340, 1545 (2013).

\bibitem{Willner2015} A. E. Willner, H. Huang, Y. Yan, Y. Ren, N. Ahmed, G. Xie, C. Bao, L. Li, Y. Cao, Z. Zhao, J. Wang, M. P. J. Lavery, M. Tur, S. Ramachandran, A. F. Molisch, N. Ashrafi, and S. Ashrafi, Optical communications using orbital angular momentum beams, Adv. Opt. Photonics 7, 66 (2015).

\bibitem{Milione2015} G. Milione, T. A. Nguyen, J. Leach, D. A. Nolan, and R. R. Alfano, Using the nonseparability of vector beams to encode information for optical communication, Opt. Lett. 40, 4887 (2015).

\bibitem{Gregg2015} P. Gregg, P. Kristensen, and S. Ramachandran, Conservation of orbital angular momentum in air-core optical fibers, Optica 2, 267 (2015).

\bibitem{Zhu2017} G. Zhu, Y. Chen, Y. Liu, Y. Zhang, and S. Yu, Characterizing a 14$\times$14 OAM mode transfer matrix of a ring-core fiber based on quadrature phase-shift interference, Opt. Lett. 42, 1257 (2017).

\bibitem{Wanga2018} A. Wang, L. Zhu, L. Wang, J. Ai, S. Chen, and J. Wang, Directly using 8.8-km conventional multi-mode fiber for 6-mode orbital angular momentum multiplexing transmission, Opt. Express 26, 10038 (2018).

\bibitem{Yao2011} A. M. Yao and M. J. Padgett, Orbital angular momentum: origins, behavior and applications, Adv. Opt. Photonics 3, 161 (2011).

\bibitem{Fickler2012} R. Fickler, R. Lapkiewicz, W. N. Plick, M. Krenn, C. Schaeff, S. Ramelow, and A. Zeilinger, Quantum Entanglement of High Angular Momenta, Science 338, 640 (2012).

\bibitem{Ding2016} D. Ding, W. Zhang, S. Shi, Z. Zhou, Y. Li, B. Shi, and G. Guo, High-dimensional entanglement between distant atomic-ensemble memories, Light Sci. Appl. 5, e16157 (2016).

\bibitem{Wang2015} X.-L. Wang, X.-D. Cai, Z.-E. Su, M.-C. Chen, D. Wu, L. Li, N.-L. Liu, C.-Y. Lu, and J.-W. Pan, Quantum teleportation of multiple degrees of freedom of a single photon, Nature 518, 516 (2015).

\bibitem{D'Ambrosio2012} V. D'Ambrosio, E. Nagali, S. P. Walborn, L. Aolita, S. Slussarenko, L. Marrucci, and F. Sciarrino, Complete experimental toolbox for alignment-free quantum communication, Nat. Commun. 3, 961 (2012).

\bibitem{Vallone2014} G. Vallone, V. D’Ambrosio, A. Sponselli, S. Slussarenko, L. Marrucci, F. Sciarrino, and P. Villoresi, Free-Space Quantum Key Distribution by Rotation-Invariant Twisted Photons, Phys. Rev. Lett. 113, 060503 (2014).

\bibitem{Bennett1984} C. H. Bennett and G. Brassard, in Proceedings of IEEE International Conference on Computers, Systems and Signal Processing (IEEE, New York, 1984), p. 175.

\bibitem{Scarani2009} V. Scarani, H. Bechmann-Pasquinucci, N. J. Cerf, M. Du$\breve{s}$ek, N. L$\ddot{u}$tkenhaus, and M. Peev, The security of practical quantum key distribution, Rev. Mod. Phys. 81, 1301 (2009).

\bibitem{Wang2005} X.-B. Wang, Phys. Rev. Lett. 94, Beating the Photon-Number-Splitting Attack in Practical Quantum Cryptography, 230503 (2005).

\bibitem{Lo2005} H.-K. Lo, X. Ma, and K. Chen, Decoy State Quantum Key Distribution, Phys. Rev. Lett. 94, 230504 (2005).

\bibitem{Lo2012} H.-K. Lo, M. Curty, and B. Qi, Measurement-Device-Independent Quantum Key Distribution, Phys. Rev. Lett. 108, 130503 (2012).

\bibitem{Sasaki2014} T. Sasaki, Y. Yamamoto, and M. Koashi, Practical quantum key distribution protocol without monitoring signal disturbance, Nature 509, 475 (2014).

\bibitem{Lucamarini2018} M. Lucamarini, Z. L. Yuan, J. F. Dynes, and A. J. Shields, Overcoming the rate-distance limit of quantum key distribution without quantum repeaters, Nature 557, 400 (2018).

\bibitem{Cerf2002} N. J. Cerf, M. Bourennane, A. Karlsson, and N. Gisin, Security of Quantum Key Distribution Using d-Level Systems, Phys. Rev. Lett. 88, 127902 (2002).

\bibitem{Sheridan2010} L. Sheridan and V. Scarani, Security proof for quantum key distribution using qudit systems Lana, Phys. Rev. A 82, 030301(R) (2010).

\bibitem{Bradler2016} K. Bradler, M. Mirhosseini, R. Fickler, A. Broadbent, and R. Boyd, Finite-key security analysis for multilevel quantum key distribution, New J. Phys. 18, 073030 (2016).

\bibitem{Nape2018a} I. Nape, E. Otte, A. Valles, C. Rosales-Guzm$\acute{a}$n, F. Cardano, C. Denz, and A. Forbes, Self-healing high-dimensional quantum key distribution using hybrid spin-orbit Bessel states, Opt. Express 26, 26946 (2018).

\bibitem{Bouchard2018a} F. Bouchard, F. Hufnagel, D. Koutny, A. Abbas, A. Sit, K. Heshami, R. Fickler, and E. Karimi, Full characterization of a high-dimensional quantum communication channel, arXiv preprint arXiv:1806.08018 (2018).

\bibitem{Cozzolino2018} D. Cozzolino, D. Bacco, B. Da Lio, K. Ingerslev, Y. Ding, K. Dalgaard, P. Kristensen, M. Galili, K. Rottwitt, S. Ramachandran, and L. K. Oxenlowe, Fiber based high-dimensional quantum communication with twisted photons, arXiv preprint arXiv:1803.10138 (2018).

\bibitem{Bouchard2018b} F. Bouchard, K. Heshami, D. England, R. Fickler, R. W. Boyd, B.-G. Englert, L. L. S$\acute{a}$nchez-Soto, and E. Karimi, Experimental investigation of quantum key distribution protocols with twisted photons, arXiv preprint arXiv:1802.05773 (2018).

\bibitem{Larocque2017} H. Larocque, J. Gagnon-Bischoff, D. Mortimer, Y. Zhang, F. Bouchard, J. Upham, V. Grillo, R. W. Boyd, and E. Karimi, Generalized optical angular momentum sorter and its application to high-dimensional quantum cryptography, Opt. Express 25, 19832 (2017).

\bibitem{Sit2017} A. Sit, F. Bouchard, R. Fickler, J. Gagnon-Bischoff, H. Larocque, K. Heshami, D. Elser, C. Peuntinger, K. Gunthner, B. Heim, C. Marquardt, G. Leuchs, R. W. Boyd, and E. Karimi, High-dimensional intracity quantum cryptography with structured photons, Optica 4, 1006 (2017).

\bibitem{Ndagano2017} B. Ndagano, I. Nape, B. Perez-Garcia, S. Scholes, R. I. Hernandez-Aranda, T. Konrad, M. P. J. Lavery, and A. Forbes, A deterministic detector for vector vortex states, Sci. Rep. 7, 13882 (2017).

\bibitem{Mirhosseini2015} M. Mirhosseini, O. S. Magana-Loaiza, M. N. O'Sullivan, B. Rodenburg, M. Malik, M. P. J. Lavery, M. J. Padgett, D. J. Gauthier, and R. W. Boyd, High-dimensional quantum cryptography with twisted light, New J. Phys. 17, 033033 (2015).

\bibitem{Mafu2013} M. Mafu, A. Dudley, S. Goyal, D. Giovannini, M. McLaren, M. J. Padgett, T. Konrad, F. Petruccione, N. Lutkenhaus, and A. Forbes, Higher-dimensional orbital-angular-momentum-based quantum key distribution with mutually unbiased bases, Phys. Rev. A - At. Mol. Opt. Phys. 88, 032305 (2013).

\bibitem{Groblacher2006} S. Gr$\ddot{o}$blacher, T. Jennewein, A. Vaziri, G. Weihs, and A. Zeilinger, Experimental quantum cryptography with qutrits, New J. Phys. 8, 75 (2006).

\bibitem{Rafsanjani2015} S. M. H. Rafsanjani, M. Mirhosseini, O. S. Maga$\tilde{n}$a-Loaiza, and R. W. Boyd, State transfer based on classical nonseparability, Phys. Rev. A 92, 023827 (2015).

\bibitem{Marrucci2006} L. Marrucci, C. Manzo, and D. Paparo, Optical Spin-to-Orbital Angular Momentum Conversion in Inhomogeneous Anisotropic Media, Phys. Rev. Lett. 96, 163905 (2006).

\bibitem{Nagali2009} E. Nagali, F. Sciarrino, F. De Martini, L. Marrucci, B. Piccirillo, E. Karimi, and E. Santamato, Quantum Information Transfer from Spin to Orbital Angular Momentum of Photons, Phys. Rev. Lett. 103, 013601 (2009).

\bibitem{Marrucci2011} L. Marrucci, E. Karimi, S. Slussarenko, B. Piccirillo, E. Santamato, E. Nagali, and F. Sciarrino, Spin-to-orbital conversion of the angular momentum of light and its classical and quantum applications, J. Opt. 13, 064001 (2011).

\bibitem{D'Ambrosio2013} V. D'Ambrosio, N. Spagnolo, L. Del Re, S. Slussarenko, Y. Li, L. C. Kwek, L. Marrucci, S. P. Walborn, L. Aolita, and F. Sciarrino, Photonic polarization gears for ultra-sensitive angular measurements, Nat. Commun. 4, 2432 (2013).

\bibitem{Ma2005} X. Ma, B. Qi, Y. Zhao, and H.-K. Lo, Practical decoy state for quantum key distribution, Phys. Rev. A 72, 012326 (2005).

\bibitem{Gottesman2004} D. Gottesman, H.-K. Lo, N. L$\ddot{u}$tkenhaus, and J. Preskill, Security of quantum key distribution with imperfect devices Daniel, Quantum Inf. Comput. 4, 325 (2004).

\bibitem{Cheny2018} Y. Chen, J. Gao, Z.-Q. Jiao, K. Sun, W.-G. Shen, L.-F. Qiao, H. Tang, X.-F. Lin, and X.-M. Jin, Mapping Twisted Light into and out of a Photonic Chip, Phys. Rev. Lett. 121, 233602 (2018).

\bibitem{Mousavi2017} S. F. Mousavi, R. Nouroozi, G. Vallone, and P. Villoresi, Integrated optical modulator manipulating the polarization and rotation handedness of Orbital Angular Momentum states, Sci. Rep. 7, 3835 (2017).

\bibitem{Mousavi2018} S. F. Mousavi, G. Vallone, P. Villoresi, and R. Nouroozi, Generation of mutually unbiased bases for 4D-QKD with structured photons via LNOI photonic wire, J. Opt. 20, 095802 (2018).

\bibitem{Gibson2004} G. Gibson, J. Courtial, M. J. Padgett, M. Vasnetsov, V. Pas'ko, S. Barnett, and S. Franke-Arnold, Free-space information transfer using light beams carrying orbital angular momentum, Opt. Express 12, 5448 (2004).

\bibitem{Leach2002} J. Leach, M. J. Padgett, S. M. Barnett, S. Franke-Arnold, and J. Courtial, Measuring the Orbital Angular Momentum of a Single Photon, Phys. Rev. Lett. 88, 257901 (2002).

\bibitem{Leach2004} J. Leach, J. Courtial, K. Skeldon, S. M. Barnett, S. Franke-Arnold, and M. J. Padgett, Interferometric Methods to Measure Orbital and Spin, or the Total Angular Momentum of a Single Photon, Phys. Rev. Lett. 92, 013601 (2004).

\bibitem{Berkhout2010} G. C. G. Berkhout, M. P. J. Lavery, J. Courtial, M. W. Beijersbergen, and M. J. Padgett, Efficient Sorting of Orbital Angular Momentum States of Light, Phys. Rev. Lett. 105, 153601 (2010).

\bibitem{O'Sullivan2012} M. N. O'Sullivan, M. Mirhosseini, M. Malik, and R. W. Boyd, Near-perfect sorting of orbital angular momentum and angular position states of light, Opt. Express 20, 24444 (2012).

\bibitem{Mirhosseini2013} M. Mirhosseini, M. Malik, Z. Shi, and R. W. Boyd, Efficient separation of the orbital angular momentum eigenstates of light, Nat. Commun. 4, 2781 (2013).

\bibitem{Wen2018} Y. Wen, I. Chremmos, Y. Chen, J. Zhu, Y. Zhang, and S. Yu, Spiral Transformation for High-Resolution and Efficient Sorting of Optical Vortex Modes, Phys. Rev. Lett. 120, 193904 (2018).

\bibitem{Nape2018b} B. Ndagano, I. Nape, M. A. Cox, C. Rosales-Guzman, and A. Forbes, Creation and detection of vector vortex modes for classical and quantum communication, J. Light. Technol. 36, 292 (2018).


\end{thebibliography}
\end{document}